\documentclass[showpacs,amsfonts,amsmath,twocolumn,floatfix,aps]{revtex4}
\usepackage{amsfonts,amsmath}
\usepackage{epsfig}

\newcommand{\beq}{\begin{equation}}  
\newcommand{\eeq}{\end{equation}}  
\newcommand{\bea}{\begin{eqnarray}}  
\newcommand{\eea}{\end{eqnarray}}  
  
\begin{document}     
\title{Lifetimes of noisy repellors}
\author{
Holger Faisst and Bruno Eckhardt
}     
\address{Fachbereich Physik, Philipps Universit\"at    
         Marburg, D-35032 Marburg, Germany}    
  
\begin{abstract}
We study the effects of additive noise on the lifetimes of 
chaotic repellors. Using first order perturbation theory we argue
that noise will increase the lifetime if the escape holes lie
in regions where the unperturbed density is higher than in the
immediate vicinity and that it decreases if the density is lower.
Numerical experiments support the qualitative conclusions 
also beyond perturbation theory.
\end{abstract}     
\maketitle

\section{Introduction}
Noise can affect the behaviour of dynamical systems in many ways.
It can induce transitions between otherwise disconnected regions
(Kramers' theory,~\cite{Haenggi90}), it changes the scaling near
bifurcations~\cite{Martin}, it gives rise to stochastic 
resonance~\cite{Haenggi98} and 
it can even change repellors into attractors~\cite{arnold83}.
Some time ago Franaszek~\cite{Franaszek91} studied the effects
of additive noise on repellors~\cite{Tel} and found that in some
cases it stabilized the dynamics, i.e.~increased the life times.
He studied this behaviour near crises and bifurcations in the 
dynamics, but the reasons for the effects on the dynamics
remained unclear. We here want to approach the problem from
the side of the attractor which
then is perturbed to become a chaotic repellor. We will 
investigate the relation between the noise effects and 
a non-uniform density in the attractor. 

The hypothesis we want to test runs as follows:
opening up the attractor into a repellor is achieved
by punching holes into the support of the attractor. Additive noise 
can push trajectories that would barely miss the holes into escape,
but can also save trajectories that would escape in the unperturbed
situation. Whether the life time increases or decreases then
depends on which process is more likely: for a uniform density
noise will kick out as many trajectories as it saves, so one
cannot expect any effect. If the unperturbed density in the hole
region is higher than in the immediate vicinity, 
more points will be saved than kicked out, and the 
lifetime should increase. If the unperturbed density in the 
hole region is lower, more trajectories will escape and 
the lifetime should be reduced. 

In section~\ref{sec_pt} we present the perturbative arguments,
followed by numerical experiments in section~\ref{sec_plm} and some 
final remarks in section~\ref{sec_fr}.

\section{Perturbation theory\label{sec_pt}}
We start with a $d$-dimensional map 
\beq
{\bf x}_{n+1} = {\bf f}({\bf x}_n)
\label{kick}
\eeq
that has a chaotic attractor. The associated Frobenius-Perron equation
for the evolution of densities $\rho({\bf x})$ is
\bea
\rho_{n+1}({\bf x}) &=& \int d{\bf y} \, 
\delta({\bf x} - {\bf f}({\bf y}))\rho_n({\bf y})\\
&=& \sum_{i} \frac{\rho_n({\bf y_i})}{|D{\bf f}({\bf y}_i)|}\,,
\eea
where $|D{\bf f}|$ is the Jacobi determinant
and the summation extends over all points ${\bf y}_i$
that map into ${\bf x}$.
Noise can be added as a Gaussian smearing of the propagator,
as in a kicked system: the time evolution splits into two parts,
the `kick' (\ref{kick}), i.e.~the application of the
deterministic map, and a free diffusive spreading. The free
diffusion on the $d$-dimensional phase space is described by 
a diffusion kernel $K_D$ that solves the appropriate free
diffusion equation with $\delta$-function initial conditions.
For instance, for free diffusion in Euclidean space the 
Fokker-Planck equation for a density $\rho$ is
\beq
\dot \rho = D \Delta \rho
\eeq
and the diffusion kernel becomes
\beq
K_D({\bf y}, {\bf x}, t) =
\frac{1}{(2\pi D t)^{d/2}} e^{-({\bf y}-{\bf x})^2/2Dt} \,.
\eeq
The combined evolution of kick and diffusion for some time 
$T$ is then described by 
\bea
\rho_{n+1}({\bf x}) &=& 
\int d{\bf z}\, K_D({\bf x},{\bf z}, T)
\int d{\bf y} \, \delta({\bf z} - {\bf f}({\bf y}))\rho_n({\bf y})\\
&=&
\int d{\bf y}\, K_D({\bf x},{\bf f}({\bf y}),T )\rho_n({\bf y})\\
&=& \int d{\bf y}\, K({\bf x}, {\bf y}) \rho_n({\bf y}) \,.
\label{evolution}
\eea

The evolution kernel $K$ in (\ref{evolution}) can be expanded
in terms of left $\langle \lambda|$ and right eigenfunctions 
$|\lambda\rangle$ with eigenvalues $\lambda$,
\beq
K = \sum_\lambda \lambda\,  |\lambda\rangle   \langle\lambda| \,.
\eeq
The existence of an invariant density on the attractor (the
Sinai-Ruelle-Bowen measure) implies the presence of
an eigenvalue $\lambda=1$ with right eigenvector $|1\rangle$,
the invariant density, and a corresponding left eigenvector
$\langle 1|=1$ because of conservation of probability. 
Since the kernel $K$ is not self adjoint left- and right-eigenvectors
are different. Simple examples show that the 
left 
eigenvectors
develop fractal features \cite{Gaspard}. With noise the 
finest scale structures are washed out, but higher levels
of the fractal hierarchy survive. 

In order to turn the attractor into a chaotic repellor we
punch holes into it. In applications the holes
appear through crises and other perturbations and appear on many scales.
For the purpose of the present analysis trajectories that enter the holes 
may be terminated, since their further evolution does not influence the 
escape rate.
Let ${\cal O}$ be the domain over which
trajectories are taken out and $P$ the elimination projection:
\beq
P({\bf x}, {\bf z}) = \delta({\bf x}-{\bf z}) \cdot
\left\{ \begin{matrix}
1 & {\bf x}\notin {\cal O}\cr
                            p & {\bf x}\in {\cal O}
\end{matrix}\right. \,.
\label{eli_pro}
\eeq
This projection depends on a parameter $p$ that will be useful in tests 
of the perturbation calculations:
Trajectories entering ${\cal O}$ continue on with
probability $p$ and are taken out with probability $1-p$.
For the holes described before we have to take $p=0$. 
The full evolution operator can then be written
\bea
& & K_P({\bf x}, {\bf y}) = 
\int d{\bf z} P({\bf x}, {\bf z}) K({\bf z}, {\bf y})\\
&=& K({\bf x}, {\bf y}) -
\int_{\cal O} 
d{\bf z} \left(
\delta({\bf x}, {\bf z}) -P({\bf x}, {\bf z})
\right)  K({\bf z}, {\bf y})\\  
&=& K({\bf x}, {\bf y}) +\alpha K_1({\bf x}, {\bf y}) \,.
\eea
The last equation now has the form of an unperturbed propagator $K$ 
plus a small perturbation $\alpha K_1$, 
where smallness is controlled by the
localization in the region ${\cal O}$ and
the rate $1-p$ with which points are taken out. $\alpha$ is a formal
parameter that helps to organize the familiar perturbation expansion
for eigenvalues and eigenvectors. The leading order result contains
as usual the diagonal matrix element of the perturbation,
\beq
\lambda \approx \lambda_0 + \alpha \langle \lambda | K_1 | \lambda \rangle\,.
\eeq
The deviations of the leading eigenvalue from $\lambda_0=1$ 
define a decay rate, $\lambda = \exp(-\gamma)$, where 
\beq
\gamma \approx -\alpha \langle \lambda | K_1 | \lambda \rangle \,.
\label{gamma}
\eeq
The systems we study here have piecewise smooth invariant densities 
and the first order correction to the decay rate has a regular 
dependence on the set ${\cal O}$ and the extraction rate $p$.
In particular, for the logistic map with holes with $p=0$ 
that is studied in 
\cite{Paar97} expression (\ref{gamma}) 
gives the smooth background. 
The simulations by Paar and Pavin~\cite{Paar97} also indicate
strong variations in escape rate near short periodic orbits 
which are connected to higher orders in perturbation theory
and the complicated spatial structures that are characteristic
for next to leading eigenvectors~\cite{Gaspard}.

\section{Piecewise linear maps 
with non-constant invariant densities\label{sec_plm}}
Let us define a family of 1-d maps $x_{n+1} = f_c(x_n)$ with
a parameter $c\in(0,0.5)$,
\beq
f_c(x) = 
\left\{
\begin{array}{cl}
x/c & 0<x\le c
\\
1-2(x-c)c/(1-2c)&  c <x \le 1/2
\\
c - (2x-1)c/(1-2c)& 1/2 <x \le 1-c  
\\
1+(x-1)/c &  1-c <x \le 1  \, .      
\end{array}
\right. 
\label{map}
\eeq
The invariant density is a solution to the Frobenius-Perron equation 
\bea
\rho_c(x)&=&\int_0^1 dy \,\delta[x-f_c(y)]\rho_c(y) \\
&=& \sum_{x=f_c(x_i)} \frac{\rho_c(x_i)}{|f_c'(x_i)|}\,,
\label{fpe}
\eea
where the sum is taken over all pre-images $x_i$ of $x$.
For (\ref{map}) the normalized invariant density is
\beq
\rho_c (x) = 
\left\{
\begin{array}{cl}
\frac{1}{4c(1-c)} & 0 < x < c \; \mbox{and}\; 1-c <x< 1
\\
\frac{1}{2(1-c)}   &  c <x< 1-c \, .        
\end{array}
\right.
\label{rho_1}
\eeq
The main results do not depend on the specific value of $c$ so that we can fix $c=0.2$ 
and drop the subscript on $f$ and $\rho$. 
\begin{figure}
\begin{center}  
\epsfig{file=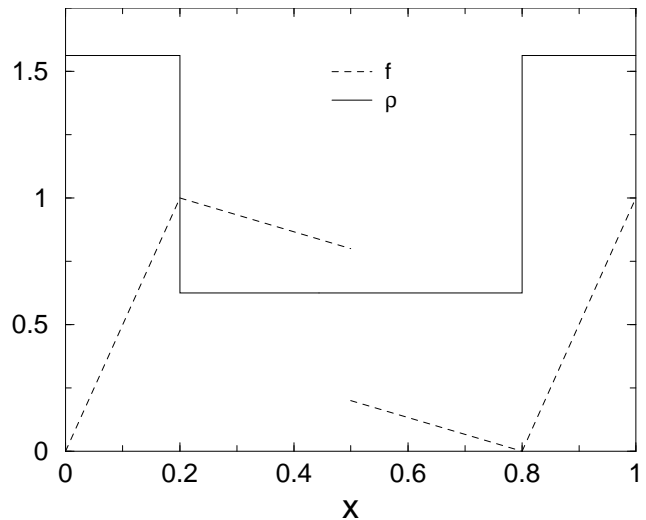,width=.47\textwidth,clip=}
\end{center}  
\caption[]{A graph of the map (\ref{map}) for $c=0.2$ (dashed) and 
its invariant density $\rho$ (continuous). 
}
\label{map_del_0.2}
\end{figure}

For the eigenvalue and eigenvector analysis we use a matrix representation of the 
density evolution operator $K$. With the help of $N$ characteristic functions $\phi_\nu$, 
\beq
\phi_\nu(x) = 
\left\{
\begin{array}{cl}
1, & (\nu-1)/N < x < \nu/N, \, \nu=1,\ldots,N
\\
0,   &  \hbox{elsewhere\, ,}        
\end{array}
\right.
\label{phi}
\eeq
densities can be expanded as $\rho= a_\nu \phi_\nu $ (summation implied) 
and the evolution operator $K$ becomes an $N \times N$ matrix. Densities are
mapped according to 
\beq
a_\nu^{(n+1)} = K_{\nu\mu}a_\mu^{(n)}\,,
\eeq
and the matrix elements $K_{\nu\mu}$ can be calculated from the images of 
the step functions: in the support of the characteristic function $\mu$
a uniformly distributed ensemble of $5\times 10^5$ points is iterated
and the probability to end up in the interval $\nu$ is then the matrix
element $K_{\nu\mu}$. The typical size of matrices used in the 
calculations is $N=3000$.

The lifetimes for the maps with holes can be obtained from the 
eigenvalues of the evolution operator after projection onto 
the remaining intervals or directly from integrations of an ensemble 
of initial conditions. We followed
$10^6$ randomly selected initial points up to a maximal cut-off lifetime of 
$10^4$ iterations. The lifetime distributions decayed exponentially and
the lifetimes estimated from this decay were within less than $1\%$ of the
eigenvalues. 

\subsection{No holes and no noise}


\begin{figure}
\begin{center}  
\epsfig{file=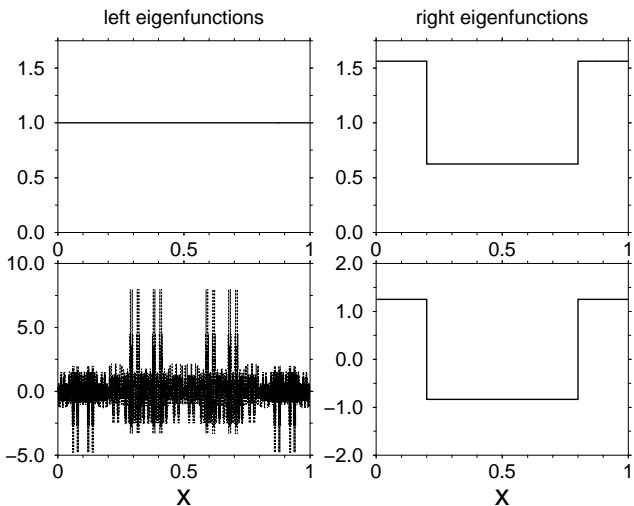,width=.47\textwidth,clip=}
\end{center}  
\caption[]{Eigenfunctions of map without noise and without holes. 
The upper row corresponds to the leading eigenvalue $\lambda_0=1$ and the 
lower one to the next to leading eigenvalue $\lambda_1=-0.6$.
}
\label{first_two_eigenfunctions}
\end{figure}

Without noise and without holes the map (\ref{map}) maps the interval
$[0,1]$ into itself. We expect one eigenvalue $1$ with right eigenvector the
invariant density (as in Fig.~\ref{map_del_0.2}), and a left eigenvector that
is constant because of conservation of probability.
The first two pairs of left and right eigenvectors are
shown in Fig.~\ref{first_two_eigenfunctions}. The next to leading eigenvalue
is $\lambda_1=-0.6$. Its left eigenvector shows the fractal structures
one expects for such maps \cite{Gaspard}. 


\subsection{With holes but without noise}

Pairs of holes of size $\epsilon$ are added symmetrically at the edges of 
the invariant density. 
`Outer' holes at $(c -\epsilon, c)$ and $(1-c, 1-c+\epsilon)$ lie within the 
high density region,  
`inner' holes at $(c, c+\epsilon)$ and $(1-c-\epsilon, 1-c)$ in the 
low density region.  

\begin{figure}
\begin{center}  
\epsfig{file=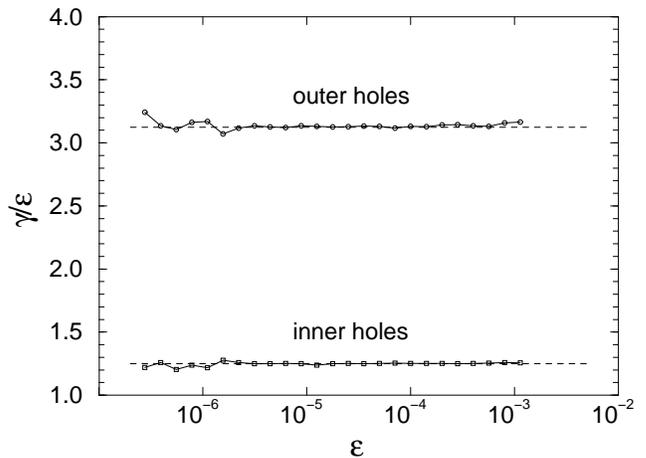,width=.47\textwidth,clip=}
\end{center}  
\caption[]{Escape rate $\gamma$ 
divided by the hole size $\epsilon$
for the map (\ref{map}) with holes but without noise.
The dashed lines at $3.125$ and $1.125$ mark the values
expected within first order perturbation theory.
}
\label{escaperate_nonoise}
\end{figure}

The perturbation theory from section~\ref{sec_pt} predicts a linear variation of escape 
rate with hole size, so that the ratio of escape rate to hole size should be
constant. This is verified in Fig.~\ref{escaperate_nonoise}.
The numerical values for the escape rate $\gamma$ agree well with the expected values,
sum of hole sizes times undisturbed density at holes, 
$\gamma_{i,o} = 2\epsilon\rho_{i,o}$
.   

\begin{figure}
\begin{center}  
\epsfig{file=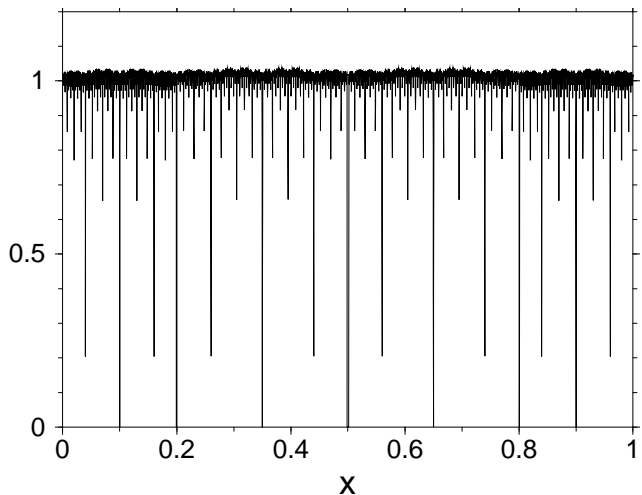,width=.47\textwidth,clip=}
\end{center}  
\caption[]{Left eigenfunction of map with outer holes of size 
$\epsilon=10^{-3}$ for the leading
eigenvalue $\lambda_0$. The hierarchy of peaks follows the pre-images of the holes.
}
\label{first_right_eigenfctn_eps_1.e-3}
\end{figure}

With holes we no longer have conservation of probability, and the left eigenvector
to the leading eigenvalue will not be constant. It develops a fractal
structure, that already for hole size $\epsilon=10^{-3}$ 
is difficult to represent numerically. The left eigenstate for 
the leading eigenvalue $\lambda_0=0.996855$ 
for the map with outer holes 
is shown in 
Fig.~\ref{first_right_eigenfctn_eps_1.e-3}.
The largest eigenvalue corresponds to an escape rate 
$\gamma_o = 3.150\times10^{-3}$ in good agreement with the values extracted 
from Fig.~\ref{escaperate_nonoise}.

\subsection{Noisy map without holes}
For the map with noise we add Gaussian distributed random numbers at
each time step,
\beq
x_{n+1} = f(x_n) + \xi_n \, ,      
\eeq
where the $\xi_n$ are independent and identically distributed according to 
\beq
p(\xi) = \frac{1}{\sqrt{\pi \sigma^2}} e^{-\xi^2/\sigma^2} \, ,      
\eeq
with $\sigma=\sqrt{2Dt}$ the amplitude of the noise.
Since it then is possible to leave the interval $[0,1]$ 
we close the system periodically by mapping points outside the interval
back in using the modulo operation.
The case with noise is the more regular one, and if the width of the
characteristic function for the hole region is a fraction of the noise 
level the vectors converge rather reliably, as a comparison
between the results for matrix sizes $N=3000$ and $N=4000$ for noise 
amplitude $10^{-3}$ showed.

\begin{figure}
\begin{center}  
\epsfig{file=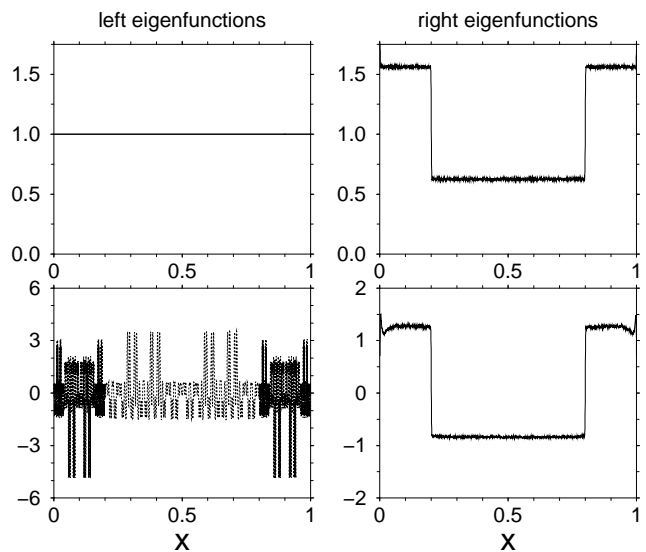,width=.47\textwidth,clip=}
\end{center}  
\caption[]{Eigenfunctions for the noisy map without holes. The noise
amplitude is $\sigma=10^{-3}$. The upper row
corresponds to the leading eigenvalue $\lambda_0=1$, the lower to the 
next to leading eigenvalue $\lambda_1=-0.6073$.
}
\label{first_two_eigenfcts_noise_1.e-3}
\end{figure}

\begin{figure}
\begin{center}  
\epsfig{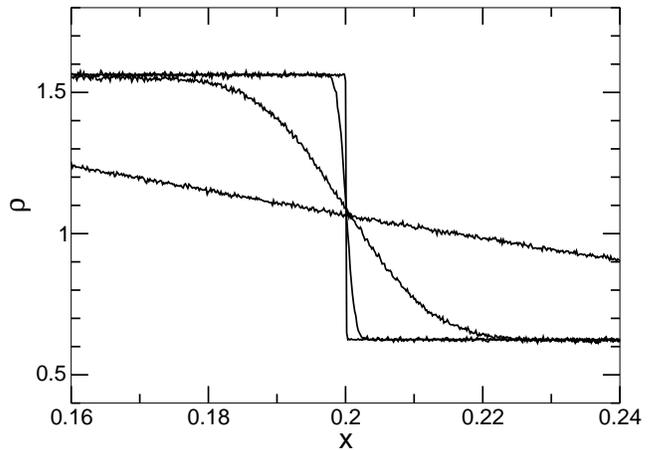}
\end{center}  
\caption[]{Magnification of the invariant density for the noisy map without holes. 
Noise amplitudes are $\sigma=10^{-5}$ (sharpest transition), $10^{-3}$, $10^{-2}$, and $10^{-1}$ (widest transition).
}
\label{smooth2}
\end{figure}

The eigenfunctions and eigenvalues in the presence of noise of amplitude
$\sigma=10^{-3}$ are shown in 
Fig.~\ref{first_two_eigenfcts_noise_1.e-3}.
The first eigenvalue remains at $1$, the first left eigenvector is uniform, but the 
step in the first right eigenvector is smoothed out. With increasing noise
amplitude this transition region becomes wider, as evidenced by the magnifications in Fig.~\ref{smooth2}.

\subsection{Noisy map with holes}
\begin{figure}
\begin{center}  
\epsfig{file=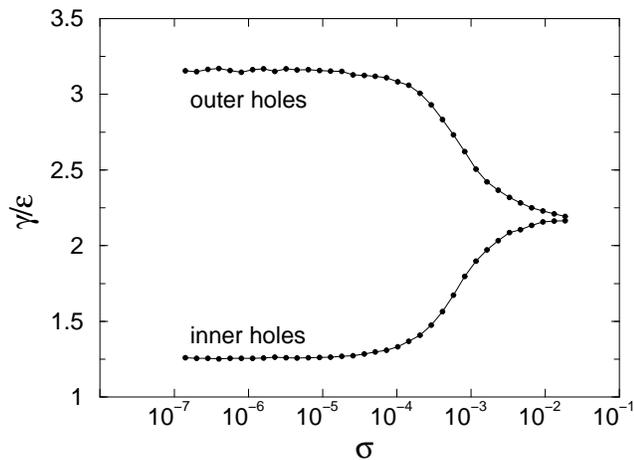,width=.47\textwidth,clip=}
\end{center}  
\caption[]{Escape rate $\gamma$ 
divided by the hole size $\epsilon = 10^{-3}$
for the noisy map 
with holes. When the holes
are in the high density region (outer holes) the escape rate decreases, and
when they are in the low density region (inner holes) it increases. When noise
level $\sigma$ and hole size are comparable there is no difference between inner and 
outer hole placement anymore.
}
\label{escaperate_eps_1.e-3}
\end{figure}

We finally come to our model for a noisy repellor, the noisy map with holes.
Fig.~\ref{escaperate_eps_1.e-3} shows for holes of size 
$\epsilon = 10^{-3}$
the change 
in the escape rates as a function of noise amplitude. 
For outer holes the escape rate decreases, for inner holes it increases, until
for a noise amplitude of about $10^{-2}$ they almost coincide.
This is exactly what one would expect from the changes in invariant density
shown in Fig.~\ref{smooth2}: the density on the upper level decreases
and the one on the lower increases with the corresponding changes in 
lifetime. 
If we had not closed the interval to a circle the loss of trajectories at the 
end of the intervals would have swamped this effect and the 
escape rate would have increased monotonically with noise level.

\begin{figure}
\begin{center}  
\epsfig{file=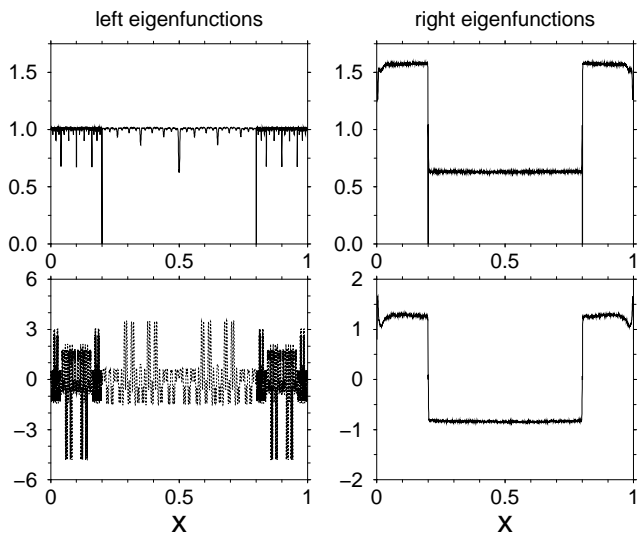,width=.47\textwidth,clip=}
\end{center}  
\caption[]{Eigenfunctions of the map with noise amplitude $\sigma=10^{-3}$ and 
with outer holes of size $\epsilon = 10^{-3}$. The eigenvalues are
$\lambda_1=0.99745$ for the upper graph and $\lambda_2=-0.61138$ for the lower one.
}
\label{first_two_eigenfcts_eps_1.e-3_noise_1.e-3}
\end{figure}

The associated eigenfunctions are shown in Fig.~\ref{first_two_eigenfcts_eps_1.e-3_noise_1.e-3} for 
outer holes of size $10^{-3}$ together with a noise amplitude 
$10^{-3}$. Compared to the noise free case
structures are smoothed out: for instance, the amplitudes of left eigenfunctions decrease considerably.

\subsection{Failure of perturbation theory}

\begin{figure}[b]
\begin{center}  
\epsfig{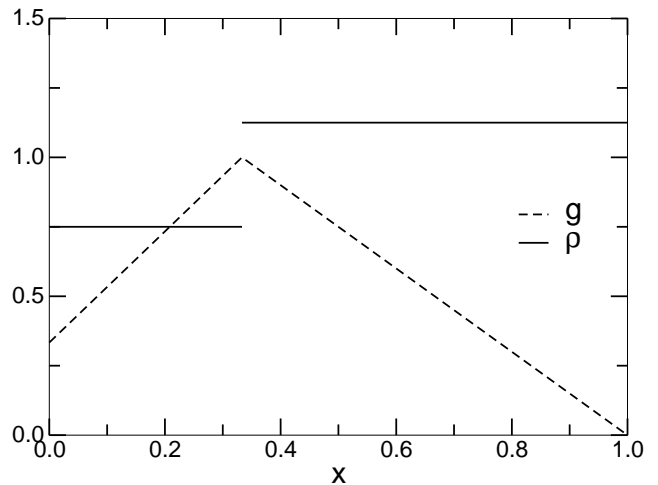}
\end{center}  
\caption[]{The map (\ref{g}) (dashed) and its invariant density (continuous).
}
\label{asymm}
\end{figure}

First order perturbation theory does not always work that well.
Consider another map, $x_{n+1}=g(x_n)$ on the interval $[0,1]$ with
\beq
g(x) = 
\left\{
\begin{array}{cl}
1/3+2x & 0<x\le 1/3
\\
3(1-x)/2 &  1/3<x \le 1
\end{array}
\right. \, .      
\label{g}
\eeq
Its invariant density is (Fig.~\ref{asymm}): 
\beq
\rho(x) = 
\left\{
\begin{array}{cl}
3/4, & 0 < x \le 1/3
\\
9/8,   &  1/3 <x \le 1        
\end{array}
\right. \, .      
\label{rho_3}
\eeq

\begin{figure}[htb]
\begin{center}  
\epsfig{file=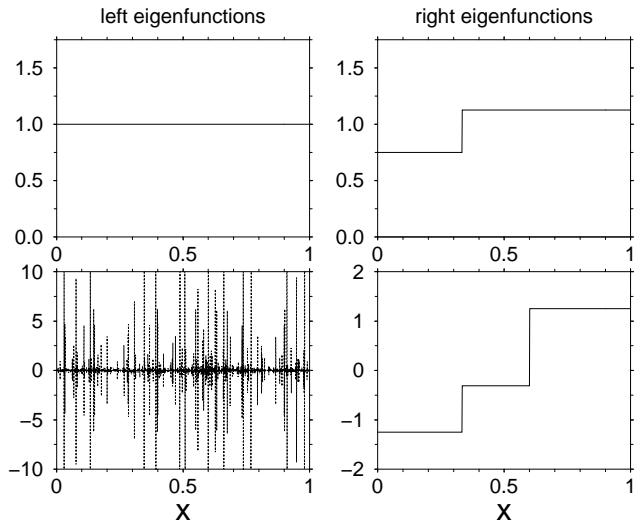,width=.47\textwidth,clip=}
\end{center}  
\caption[]{Eigenfunctions for the leading eigenvalue $\lambda_0=1$ and the next to leading eigenvalue $\lambda_1=-2/3$
for map (\ref{g}). }
\label{eigenfctns_asymm_map_noholes}
\end{figure}

\begin{figure}
\begin{center}  
\epsfig{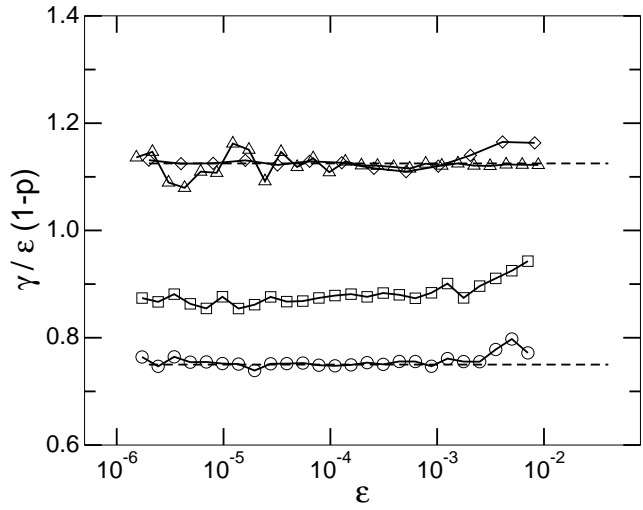}
\end{center}  
\caption[]{Escape rates $\gamma$ 
divided by hole size $\epsilon$ and by $(1-p)$
for the map (\ref{g}) vs.~hole size for two placements
of the hole and two extraction rates $p$. 
The dashed horizontal lines mark the theoretical values $9/8$ and $3/4$. For the left hole the escape rate is in
agreement with perturbation theory (circles, for p=0). For the opening to the
right at $p=0$ (squares) the pre-factor does not agree with perturbation theory.
For $p=0.99$ the values move up to the first order perturbation theory
result (triangles). If the position of the opening is moved away from the
critical point $x=1/3$, e.g.~to the interval $[0.35,0.35+\epsilon]$, then
the perturbative results is also obtained for $p=0$ (diamonds).
}
\label{escaperates_vs_eps_nonoise}
\end{figure}

Eigenfunctions for the two leading eigenvalues are shown in 
Fig.~\ref{eigenfctns_asymm_map_noholes}.
Again we take a `left' hole at $[1/3-\epsilon,1/3]$ in the low density region
and a `right' hole at $[1/3,1/3+\epsilon]$ in the high density region.
First order perturbation theory predicts the decay rates $\gamma$ 
as function of hole size $\epsilon$ to be
\bea 
\gamma_r &=& \epsilon \rho_r = 9/8 \epsilon \\
\gamma_l &=& \epsilon \rho_l = 3/4 \epsilon \, . 
\eea 
The numerical values extracted from Fig.~\ref{escaperates_vs_eps_nonoise} 
are 
\bea 
\gamma_r &=& 
0.87 \epsilon \\
\gamma_l &=& 
0.75 \epsilon \, .  
\eea 
In agreement with perturbation theory the escape rate is proportional
to the hole size. 
The pre-factors from the two calculations agree for the left hole in the low, but
disagree for the right one in the high density region. 
The deviation becomes smaller when the holes are
partially closed, i.e., the parameter $p$ in (\ref{eli_pro}) is set to $0.99$.
Then the perturbative and numerical escape rates are in better 
agreement. This failure of the perturbative estimate seems to be
closely connected with the presence of a periodic orbit on the 
border of the interval. If the opening is shifted to slightly
larger values then the slope agrees again with the perturbative results.

\section{Final remarks\label{sec_fr}}
Within the simple models studied here the hypothesis that the variation of 
lifetimes can be related to inhomogeneities in the invariant density of the 
unperturbed attractor could be confirmed. Such inhomogeneities are most 
pronounced near bifurcations and crises \cite{Grebogi83}, 
as in the work of Franaszek \cite{Franaszek91}.
The placement of holes from the outside is less artifical than it may seem. In 
the case of riddling bifurcations \cite{riddles} line attractors are broken up by
holes that appear near periodic points that are no longer transversally stable, 
and both position and widths can
be controlled externally. Thus the observations discussed here have
some bearing on the effects of noise on the lifetimes in riddled attractors.

Support from the Deutsche Forschungsgemeinschaft is greatfully acknowledged.



\end{document}